\documentclass[lettersize,journal]{IEEEtran}

\usepackage{tikz}
\usepackage{amsmath}
\usepackage{array}
\usepackage{url}
\usepackage{float}
\usepackage{caption}
\usepackage{multirow}
\usepackage{booktabs}
\usepackage{makecell}
\usepackage{graphicx}
\usepackage[ruled,vlined,linesnumbered,boxed]{algorithm2e}
\usepackage{stfloats}
\usepackage{textcomp}
 \usepackage{amsfonts}
\usepackage{threeparttable}
\usepackage{amssymb} 
\usepackage{mathrsfs}
\usepackage[mathscr]{euscript}
\usepackage{enumitem}
\usepackage[colorlinks=true, allcolors=blue]{hyperref}
\usepackage{subcaption} 

\newcommand{\attack}{{\sc Write+Sync}}

\begin{document}

\title{\textsc{Write+Sync}: Software Cache Write Covert Channels Exploiting Memory-disk Synchronization}

\author{Congcong~Chen,
Jinhua Cui,
Gang~Qu,~\IEEEmembership{Fellow,~IEEE},
and Jiliang Zhang,~\IEEEmembership{Senior Member,~IEEE}
\thanks{This work was supported by the National Natural Science Foundation of China under Grants No. 62122023 and U20A20202, the Natural Science Foundation of Hunan Province (Grant No. 2023JJ40160), the Natural Science Foundation of Changsha City (Grant No. kq2208212).
(\textit{Corresponding author: Jiliang Zhang})}
\thanks{C. Chen, J. Cui and J. Zhang are with Hunan University, Changsha 410082, China (e-mail: zhangjiliang@hnu.edu.cn).}

\thanks{G. Qu is with University of Maryland, College Park, MD 20742 USA.}}


\maketitle

\begin{abstract}
Memory-disk synchronization is a critical technology for ensuring data correctness, integrity, and security, especially in systems that handle sensitive information like financial transactions and medical records. We propose {\attack}, a group of attacks that exploit the memory-disk synchronization primitives. {\attack} works by subtly varying the timing of synchronization on a software cache (i.e., the write buffer), offering two advantages: 1) implemented purely in software, enabling deployment on any hardware devices; 2) resilient against existing countermeasures. We present the principles of {\attack} through the implementation of two write covert channel protocols, using either a single file or page, and introduce three enhanced strategies that utilize multiple files and pages. The feasibility of these channels is demonstrated in both cross-process and cross-sandbox scenarios across diverse operating systems (OSes). Experimental results show that, the average rate can reach 2.036 Kb/s (with a peak rate of 14.762 Kb/s) and the error rate is 0\% on Linux; when running on macOS, the average rate achieves 10.211 Kb/s (with a peak rate of 253.022 Kb/s) and the error rate is 0.004\%. To show its security implications, we evaluate it using two case studies--website fingerprinting and performance degradation attacks. To the best of our knowledge, {\attack} is the first high-speed write covert channel for software cache.
\end{abstract}

\section{Introduction}

In modern computer systems, the integration of various components has significantly boosted performance and efficiency. The most representative of these is caching. Caches leverage temporal and spatial locality to retrieve or prefetch frequently accessed data into fast storage closer to processing units. This reduces the frequency of memory and disk accesses, thus enhancing computing performance. In general, caches are shared among multiple processes, bringing substantial benefits in terms of efficiency and cost. However, this sharing of resources opens up the door to covert channel attacks \cite{b1}. Specifically, an attacker can bypass system security isolation by regulating the state of shared caches on one end and measuring state changes on the other end to covertly transmit confidential information \cite{b46}. These channels have been shown to be effective in the cross-process, cross-virtual machine (VM), and cross-sandbox scenarios, and more recently are utilized in combination with transient execution \cite{b0} to launch practical attacks \cite{b21,b22}.

In recent decades, numerous cache covert channels have been proposed, as outlined in Table \ref{tab1}. Most of these channels target hardware caches due to their nature in terms of high transmission rates (TRs). However, such attacks require sophisticated reverse engineering, and with increased awareness, various effective mitigation measures have been proposed \cite{b10,b11,b23,b25}. With these mitigations in place, potentially effective cache covert channels may be to leverage software caches outside the CPU. Goethem et al. \cite{b12} demonstrated a timing attack based on browser cache, exploiting the time taken to download network files or parse multimedia resources to leak information. Gruss et al. \cite{b9} utilized a set of operating system calls (e.g., \texttt{mincore}) to derive page cache information, achieving covert channels without timers. These channels leverage the time difference or feedback information of data access (loads) between cache hits and misses. In contrast, we propose a group of new covert channels based on a software cache (as positioned in the lower right entry of Table 1) that exploit the time differences between writing-back cached or uncached data.

\begin{table}[tbp]
\setlength{\abovecaptionskip}{0.2cm}
\setlength{\belowcaptionskip}{-0.2cm}
\small
\renewcommand{\arraystretch}{1.1}
\caption{The taxonomy of cache covert channel}
\label{tab1}
\centering
\begin{tabular}{|m{1.3cm}<{\centering} || m{3.8cm}<{\centering} || m{2.1cm}<{\centering}|}
\hline
\textbf{Type} & \textbf{Load Channel} & \textbf{Write Channel} \\
\hline
\textbf{Hardware cache} & LLC \cite{b3,b4,b5,b6}, Cache bank \cite{b26}, Cache directories \cite{b27}, Coherence \cite{b28}, LRU states \cite{b29} & WB \cite{b7}, WRITE+WRITE \cite{b45} \\
\hline
\textbf{Software cache} & Page cache \cite{b9}, Browser cache \cite{b12} & {\attack} (this work) \\
\hline
\end{tabular}
\end{table}

The new covert channels are based on two techniques that are employed on modern OSes: (1) the delayed write, which temporarily stores data to be written to disk in a write buffer. The buffer is only discharged to the output queue for the actual I/O operation when it is full or needs to be replaced, thus improving I/O efficiency; and (2) the synchronous system calls that are utilized to promptly update file contents to disk. Such persistent writes can form some secure data synchronization points for essential applications. 

Although the aforementioned two techniques are beneficial for performance and security respectively, we discover that their combination could pose security risks by modulating the time difference on a software cache --- write buffer, which is situated in the OS kernel and shared among different processes. An attacker can establish a covert channel as long as she is able to initialize and change the state of write buffers in a way that results in time variation. We demonstrate this through five synchronous system calls --- \texttt{sync}, \texttt{fsync}, \texttt{fdatasync}, \texttt{msync}, and \texttt{fcntl} on different OSes --- all of which can be used to construct covert channels. We develop two covert channel protocols, utilizing single file and page respectively, and propose three optimization strategies. We also demonstrate the capability of such covert channels in a sandbox with almost unaffected transmission rates. Like DRAMA \cite{b14} and page cache \cite{b9} attacks, our channels can work across cores and CPUs. Finally, we present several countermeasures that can reduce the performance of the proposed channels and serve as guidelines for developers.

Our contributions can be summarized as follows:
\begin{itemize}
    \item A group of new software cache write covert channels. We scrutinize the security implications of synchronous system calls and discover a vulnerability in their implementation: \textbf{can work in read-only mode}. In light of this, we design {\attack} attacks utilizing only a single shared file or page.
    \item Three optimization strategies. We use multiple files and pages to devise three optimizations, which can effectively increase the TR of {\attack} from about 400 b/s to 2.036 Kb/s with a measured bit error rate (BER) of 0\% on Linux.
    \item A detailed performance evaluation. We conduct a series of evaluations for the proposed channels, including TR, BER, standard deviation (SD), and standard error (SE). Additionally, we demonstrate the performance of {\attack} in a cross-sandbox scenario and on various OSes. The resulting channels achieve virtually unaffected performance in sandboxes and an average TR of 10.211 Kb/s with a BER of 0.004\% on macOS.
    \item Two practical attacks and three potential mitigation strategies. We demonstrate website fingerprinting and performance degradation attacks, and propose three possible countermeasures against this type of attack.
\end{itemize}

The remainder of this paper is organized as follows. Section II presents background information on timing covert channels, write-back cache, and memory-disk synchronization. Section III covers our threat model, the design of {\attack} and two communication protocols using a file and page. Section IV details three optimization strategies and compares their advantages and disadvantages. Section V explains the synchronization in communication. Section VI evaluates the performance of {\attack} in different scenarios and compares it with published work. Section VII demonstrates practical attacks and Section VIII discusses possible countermeasures. Related work is surveyed in Section IX, and we conclude in Section X.

\section{Background}


\subsection{Timing Covert Channel}
The timing covert channel utilizes the time variation of shared resources to transmit information between two entities, thereby evading detection or surveillance. According to the type of shared resources, the timing covert channel can be divided into software and hardware channels. The software channel utilizes time differences caused by changes in the state of shared system resources. For instance, page cache attack \cite{b9} detects whether the victim has loaded a particular memory page via a system call without relying on timing measurements. Gruss et al. \cite{b19} exploit timing differences in write accesses on deduplicated pages to leak the victim's memory accesses. The hardware channel exploits hardware design flaws or special hardware designs to transmit information. The available hardware resources include cache \cite{b2,b4}, value predictors \cite{b15}, execution ports \cite{b16,b17}, memory bus \cite{b18}, branch predictors \cite{b30,b31,b40}, and many more.

Typically, three conditions are required to build a covert channel \cite{b30}: 1) the Spy sets an initial state of shared resources; 2) the Trojan can change the state; 3) the Spy is able to detect the state changes through time measurement. Taking Flush+Reload \cite{b2} as an example, the Spy first utilizes the \texttt{CLFLUSH} instruction to flush the target cache line ($Flush$), followed by a waiting period. During this interval, the Trojan chooses to load or not load the target data based on the secret transmitted. Finally, the Spy re-accesses the target data and measures the reload time ($Reload$). A low reload latency indicates that the Trojan has accessed the target data (cache hit), indicating the transmission of ``1''. Otherwise, a high latency (cache miss) indicates that the data transmitted is ``0''.

\subsection{Write-back Cache}
The file system of modern computers adopts a caching mechanism known as the write-back cache \cite{b44}, which allows for efficient data storage and retrieval. With this mechanism, when functions such as \texttt{fwrite} and \texttt{write} invoke write syscall with the \texttt{O\_SYNC} flag, the written data is temporarily stored / cached in a write buffer instead of the permanent storage. A real write to the disk storage happens only when certain conditions are met, such as the buffer filled up, timeout, or the execution of a synchronization operation. This method is called delayed / lazy write, which can reduce the cost of frequent disk access, greatly improving the performance and efficiency of the system.

\begin{figure}[t]
\centering
\includegraphics[width=0.9\linewidth]{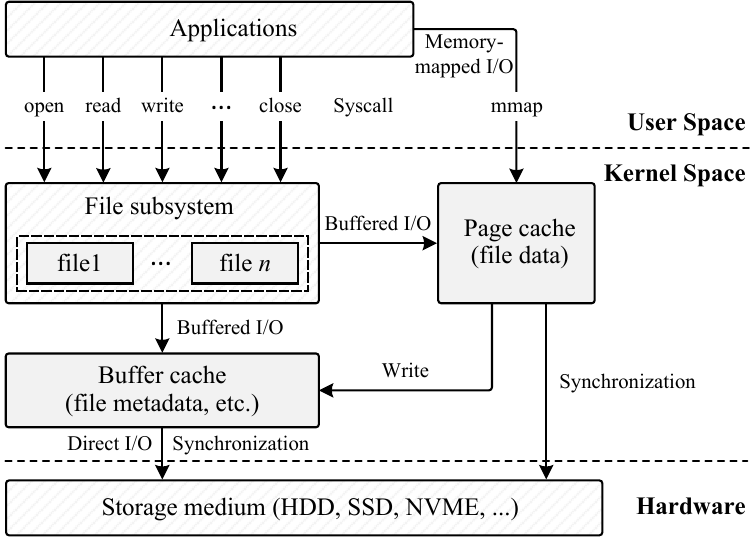}
\DeclareGraphicsExtensions.
\caption{Diagram of the write buffer system}
\label{fig1}
\end{figure}

\begin{figure*}[hbp]
\centering
\includegraphics[width=0.85\linewidth]{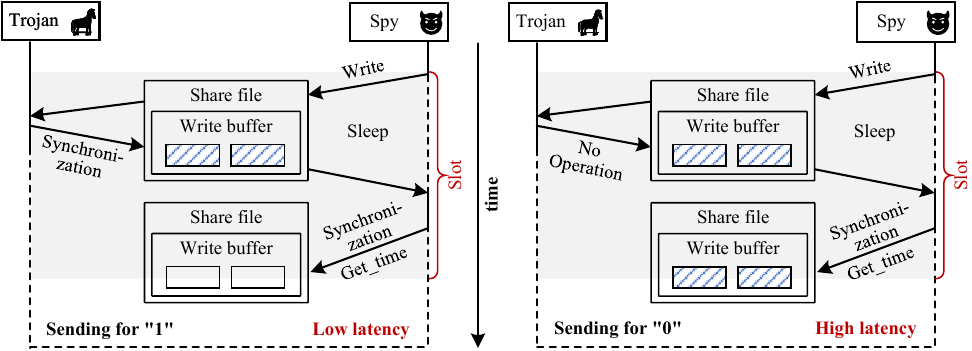}
\caption{Overview of our covert channels. The blue shaded area is cached and the blank area is uncached.}
\label{fig2}
\end{figure*}

\subsection{Memory-disk Synchronization Mechanism}
Delayed writing reduces the number of disk reads and writes, but it slows down the rate for the cases where the file contents are updated. This means that the written file data is not synchronized exactly with the external storage. Although the time interval between such desynchronization is narrow (perhaps only a few or tens of seconds depending on the data size and the state of write buffer), a power failure or system crash during the interval may result in loss of the written data. For this purpose, POSIX-compliant OSes introduce some synchronization guarantees, such as \texttt{sync}, \texttt{fsync}, \texttt{fdatasync}, \texttt{msync}, and \texttt{fcntl}. Specifically, \texttt{sync} flushes all modified block buffers to disk; \texttt{fsync} flushes all data and metadata (e.g., file size, access time) of a specified file to disk; \texttt{fdatasync} only flushes the data portion of a file, excluding metadata; \texttt{msync} synchronizes the data of a specified memory region to disk; \texttt{fcntl} with the F\_FULLFSYNC parameter is similar to the \texttt{fsync} function, but it waits for disk operations to complete before returning, rather than merely ensuring that the data has been written to the kernel buffer. These functions offer a base for memory-disk synchronization.

\section{\textsc{Write+Sync} Design}
The design of the file write buffer system in modern OSes is depicted in Fig. \ref{fig1}. When a user invokes the \texttt{write} function, the file subsystem temporarily stores the data to be written in the kernel's write buffer. This buffer comprises the page cache and buffer cache, dedicated to storing file data and metadata, respectively. Subsequently, the \texttt{write} function can promptly return to enhance I/O efficiency. The data is only written back to the disk when the write buffer is filled up or a synchronization operation is executed. However, this design introduces security implications, as elaborated in this section.

\subsection{Threat Model}

In {\attack}, we assume that the attacker has the following capabilities: 1) the attacker can control over two processes, i.e., a Trojan and a Spy; 2) the Trojan has acquired a secret through certain means. It could be an application downloaded from untrusted sources or disguised by the attacker as a benign application to deceive users into installation. Note that the Trojan cannot directly transmit sensitive data externally due to a lack of sufficient permissions or being explicitly prohibited from direct communication by isolation technologies such as sandboxing. This is prevalent in various real-world scenarios, particularly in multi-tenant cloud environments \cite{b48}; 3) the Spy can share files with the Trojan, like covert channels utilizing file locks \cite{b33,b49}. These shared files can be created by the Spy in directories accessible to the Trojan or pre-agreed by two parties in advance. Additionally, the Spy can write to the target files and is able to communicate with external networks.
All of these assumptions align with existing threat models of covert channels \cite{b2,b9}, except assumption 3). However, it is not unreasonable because the Spy is located in an external system environment where normal operations on OS resources are unrestricted. Significantly, the proposed write covert channel is purely software-based and does not require reverse engineering of specific hardware structures.

In an actual attack (see Section VII), the threat model differs from the above. The attacker only controls one process to reveal the victim's behavior. In the website fingerprinting attack, it is assumed that the attacker has access to certain log files related to website activity, which themselves do not contain sensitive information or may be encrypted. By employing synchronous system calls to monitor the write operations on these log files, the attacker can analyze users' browsing behaviors. Such log files are commonplace, especially in databases. In the performance degradation attack, the attacker utilizes synchronization to affect the performance of the target machine, thus launching a DoS-like attack \cite{b54}. Alternatively, consider a scenario where the attacker knows critical performance nodes of the victim's workload and is able to selectively induce performance degradation by synchronization functions.

\subsection{Overview}
The essence of building cache write covert channels is that the attacker has the capability to distinguish whether the data is in the write buffer or not. This can be achieved through a synchronization invocation and measuring its execution time. 
Based on this, we craft a write covert channel, as shown in Fig. \ref{fig2}. The covert channel we designed consists of three steps: 1) \textit{state initialization.} The Spy process first writes to a shared file, and the write operation returns immediately after the written data enters into the write buffer due to the delayed write (\textsc{Write}). At this point, the write buffer contains data, which we note as a cached state; 2) \textit{Trojan execution.} The Trojan process can successfully invoke the synchronization function even if its access mode to the target file is read-only. Thus, according to the secret value to be sent, the Trojan can choose whether to perform a synchronization operation (\textsc{Sync}). We assume that the Trojan synchronizes the buffered data to disk when sending ``1'', at which point the write buffer goes into an uncached state. Moreover, the Trojan does nothing when sending ``0'', and at that moment the write buffer remains cached; 3) \textit{Spy measurement.} The Spy performs synchronization on the write buffer and measures its time (\textsc{Sync}). If the Trojan has synchronized the write buffer previously (uncached), the measurement of the Spy yields a low latency. Otherwise, the Spy gets a high latency if the Trojan does nothing (cached).

The memory-disk synchronization primitives we use include \texttt{sync}, \texttt{fsync}, \texttt{fdatasync} and \texttt{msync}. According to the object types of synchronization, we classify them into synchronized files and synchronized pages. Among them, \texttt{sync}, \texttt{fsync}, and \texttt{fdatasync} are synchronized files, while \texttt{msync} is synchronized pages. In the following, we take \texttt{fdatasync} and \texttt{msync} as examples to describe the protocols of our covert channels. The protocols for other synchronization primitives are similar, so we omit them. 

\setlength{\textfloatsep}{5pt}
\begin{algorithm}[t]
\renewcommand{\algorithmcfname}{Protocol}
\DontPrintSemicolon
\caption{Single file channel with \texttt{fdatasync}}
\label{Pro1}
\KwIn{$secret[len]$, $start$, $end$, $temp$}
\KwOut{$recv[len]$}
\textbf{Sender operations:} \;
$fd$ = open(``file'', O\_RDONLY);   // Open target file \;
\For{$i\leftarrow 0$ \KwTo $len$}{
    $start$ = get\_time(); \;
    \If {$secret[i]$ == 1}{ 
        \texttt{fdatasync}($fd$);   // Synchronize file to disk \;
    } 
    \Else { 
        no operation; \;
    }
    \texttt{slotwait}($start$, slot);   // Wait for an interval \;
}
\BlankLine
\textbf{Receiver Operations:}

$fd$ = open(``file'', O\_RDWR);  \;
\For{$i\leftarrow 0$ \KwTo $len$}{
    $temp$ = get\_time(); \;
    \texttt{write}($fd$); \;
    usleep();  // Wait for the sender code to execute \;
    $start$ = get\_time(); \;
    \texttt{fdatasync}($fd$); // Synchronize file to disk \;
    $end$ = get\_time() - $start$; \;
    \If{$end < thresold$} { 
        $recv[i]$ = 1; \;  
    } 
    \Else { 
        $recv[i]$ = 0; \;
    }
    \texttt{slotwait}($temp$, slot); // Wait for an interval \;
}
\end{algorithm}

\subsection{Synchronized File}
We design a covert channel using a single file, as illustrated in Protocol \ref{Pro1}. This channel leverages the \texttt{fdatasync}, which writes all modified data blocks of a file back to the disk. To begin, the receiver initiates and writes data to the target file, and the value written can be any 1-byte content. Thanks to the delayed write, the written data is temporarily stored in the write buffer and retained for a period of time. The receiver then waits for the sender to execute, and the sender decides whether or not to call the \texttt{fdatasync} based on the secret value to be transferred. It is assumed that the sender invokes \texttt{fdatasync} when sending a ``1'' and does nothing when sending a ``0''. The receiver then calls the \texttt{fdatasync} and measures its execution time. We pre-set a threshold that lies between the time of a buffer hit and miss. If the time measured is less than the threshold, it indicates that the sender has already synchronized the dirty data from the write buffer to the disk. Thus, the receiver can infer that the sent bit is ``1''. Otherwise, if it exceeds the threshold, the receiver may deduce that the sent bit is ``0''.

It should be noted that this protocol employs a slot interval to ensure the time of each sent and received bit is equal, which is also reflected in the gray shaded area in Fig. \ref{fig2}. Concretely, upon completion of a round of code execution within the designated slot time, both the sender and receiver should wait for each other before transmitting the next bit. Following the slot constraint, the sender and receiver can maintain synchronization throughout the transmission process. This means that the sender only transmits a bit before the receiver has received one, thereby ensuring the interactive consistency of both entities. Such synchronization is referred to as inter-code synchronization, which will be detailed in Section VI.

\subsection{Synchronized Page}

The data size for each synchronization in the Protocol 1 contains all modified blocks within a file, so the granularity of synchronization is file size. However, this granularity may not be fine-grained enough if an attacker wants to monitor multiple modifications made to the same file. Therefore, we design the Protocol 2, which curtails the size of the synchronized object to the page level, thus reducing resource consumption and increasing the TR. 

In Protocol 2, we use the \texttt{msync} to flush the mapped memory segment of a specified-length of file back to disk, thus constructing a covert channel at the page level. Although the length is user-specified and can be any integer, the system aligns it with the page boundary by default. The entire protocol design is similar to Protocol 1, but we need to utilize \texttt{mmap} for page mapping and replace \texttt{fdatasync} with \texttt{msync}. Specifically, both the sender and the receiver invoke \texttt{mmap} to map the target page from a file. However, the difference is that the sender can only establish a read-only mapping of the target page, whereas the receiver is able to read and write to it. To initialize the state of write buffer, the receiver first writes data to the target page and then waits for the sender to execute. Based on the secret value to be transmitted, the sender determines whether the \texttt{msync} should be called. If it is the case, the sender needs to specify a synchronization length and set the \texttt{MS\_SYNC} flag. At last, the receiver invokes \texttt{msync} and measures its execution time. By analyzing the time variation, the receiver can deduce whether a ``1'' or ``0'' was transmitted by the sender. Notably, each bit in this protocol is assigned with a specific slot time to ensure inter-code synchronization.

\subsection{Analysis}

The aforementioned protocols offer an effective solution to the problem of inter-code synchronization. The solution is widely used in cache covert channels \cite{b2,b3,b31}, but it includes some requirements that may affect the performance of our channel. We hypothesize that the time of the synchronous system call is $t_b$ when the write buffer is cached, and that is $t_u$ when the write buffer is uncached. In our experiments, $t_b$ is much larger than $t_u$. By analyzing Protocols 1 and 2, we can reveal that both protocols have two necessary requirements:

\textbf{R1:} the sender and receiver communicate synchronously, so the receiver has to wait for the sender to complete operations. The waiting time should be large enough for dealing with different situations. For example, when the sender transmits a ``1'', the synchronization operation (e.g., \texttt{msync} call) takes a time of $t_b$, whereas the sender will do nothing when sending a ``0'' with negligible time. However, it needs to wait for a time interval of $t_b$ for the receiver to ensure that the sender completes all operations.

\textbf{R2:} each transmission requires waiting for a pre-calculated time window, referred to as a slot, to realize inter-code synchronization. The sender and receiver take different amounts of time to transmit different bits. For instance, when transmitting a ``1'', the time of synchronization for the sender is $t_b$, and that for the receiver is $t_u$. On the other hand, when sending a ``0'', the sender does nothing while the receiver calls the synchronization for a time of $t_b$. To synchronize each bit transmitted between the sender and the receiver, we need to set the slot to the maximum execution time, i.e., 2$t_b$. Note that the execution time of other codes is omitted.

These requirements could lead to long idle waiting for the sender and receiver, thereby reducing the TR of {\attack}. In addition, waiting for a slot interval per bit may affect the BER. For instance, if the slot is not large enough or the synchronization takes much time in a slot because of system blocking, the execution of the sender or receiver may exceed the slot, resulting in a corresponding bit error. Even worse, the error may be passed to the next bits. Therefore, we need to consider some optimization strategies to decrease or eliminate the effect of these requirements.

\section{Optimization Strategies}

In the previous section, we introduced two covert channel protocols that utilize a single file and page. However, a process can typically control over multiple files and pages in real-world applications. Hence, we develop three optimization strategies that leverage multiple files or pages. These strategies can reduce the above requirements and thus bring benefits in terms of TR and BER. In the following sections, we will elaborate on these strategies.

\subsection{Optimization 1: Multi-bit Encoding}

Multi-bit encoding is a coding method that enables the simultaneous transmission of multiple bits through varying levels of temporal differences. Unlike the multi-bit encoding scheme employed by hardware caches \cite{b7}, {\attack} cannot leverage time differences across different levels to encode multiple bits. This is because each file can only be synchronized separately even if multiple files are modified. Furthermore, the increase in synchronization delay resulting from modifying multiple pages is negligible, rendering it challenging for the receiver to discern such a delay.

To address the above issue, we design a multi-bit encoding scheme by different files rather than different timing levels, as shown in Optimization \ref{alg1}. Concretely, it is implemented using 4 files (No. 0, 1, 2, 3) to send two bits simultaneously. The sender groups the secret data into a batch of 2 bits, e.g., ``00'', ``01'', ``10'', and ``11''. If the secret value is ``00'', the sender synchronizes the 0th file, whereas if the secret value is ``01'', the sender synchronizes the 1th file, and so on. The sender selects one of the files to synchronize based on the secret value, so its execution time is $t_b$. In contrast, the receiver must synchronize and measure all 4 files for each received bit. Fortunately, this process can be performed concurrently, as the 4 files are not accessed in order of priority. Thus, we can use 4 threads to synchronize each of the 4 files, respectively. To reduce the time of thread creation and release, we used the thread pool technique \cite{b41} that initializes 4 threads at the beginning of our program. By this means, the execution time for the receiver would be approximately $t_b$ ideally, and the time consumed when transferring two bits is 2$t_b$. Therefore, multi-bit coding can improve the TR of our channel.

\begin{algorithm}[t]
\renewcommand{\thealgocf}{1}
\renewcommand{\algorithmcfname}{Optimization}
\DontPrintSemicolon
\caption{Multi-bit encoding with \texttt{fdatasync}}
\label{alg1}
\KwIn{$secret[len]$, $start$, $t$, $fd[4]$}
\KwOut{$recv[len]$}
\textbf{Sender operations:} \;
\For{$i\leftarrow 0$ \KwTo $len$}{
    $start$ = get\_time(); \;    
    \texttt{fdatasync}($fd[secret[i]]$); // Synchronize file$_i$ \;   
    \texttt{slotwait}($start$, slot); // Wait for an interval \;
}
\BlankLine
\textbf{Receiver operations:} \;
thpool = threadpool\_init(4);  // Initialize 4 threads \;
\For{$j\leftarrow 0$ \KwTo $len$}{
    $start$ = get\_time(); \; 
    write to 4 files; \;
    usleep(); // Wait for the sender to execute \;
    \For{$i\leftarrow 0$ \KwTo $4$}{
        $t$ = thpool\_add\_work(thpool, task, $i$); // Synchronize file$_i$ and measure time
    }
    \If{$t < thresold$} { 
        $recv[j]$ = 1; \;  
    } 
    \Else { 
        $recv[j]$ = 0; \;
    }
    \texttt{slotwait}($start$, slot); // Wait for an interval \;
}
\end{algorithm}

\subsection{Optimization 2: Asynchronous Communication}

Intuitively, Optimization 1 can use 8 or more files and threads to encode more bits at a time, but exponentially increasing resource consumption and larger interference come with that. Therefore, this limits the TR of such an approach to some extent. We propose a second option--asynchronous communication, in which the sender can continuously send messages or perform tasks without waiting for the execution of the receiver, and the receiver can process the messages at its own pace instead of responding immediately. \cite{b5} proposed the initial asynchronous communication scheme for cache covert channels, but attackers rely on a shared array larger than the LLC capacity to cause cache thrashing due to the impact of cache replacement policies. Moreover, the gap between the sender and receiver should be small to avoid interference. In contrast, {\attack} is not encumbered by this cache replacement and gap since the size of the software cache can be dynamically adjusted. Therefore, our channel can readily adapt to this type of communication. 

An important challenge for asynchronous communication is how to ensure a bounded gap between the sender and receiver. To address this, we have designed two schemes. The first scheme is to reserve a slot for each transmitted bit, which ensures that the sender is always ahead of the receiver. Concretely, the sender and receiver execute simultaneously, but the receiver waits for a slot in the initial execution. When the receiver synchronizes the 0th file, the sender is operating on the 1th file. Thus, the receiver need not to wait for the sender during subsequent execution. This scheme enables asynchronous execution between sender and receiver, which eliminates the need for \textbf{R1}. However, a desired TR cannot be achieved due to the unsolved \textbf{R2}. 


The second scheme is to use the pseudo-random encoding method \cite{b5}. Specifically, the sender uses a pseudo-random number generator (PRNG) to encode the data to be sent. For each payload bit (PB-$i$), the transmitted bit (TB-$i$) is TB-$i$ = PB-$i$ $\oplus$ PRNG-$i$. The receiver decodes the received data and uses the PRNG to recover the sent data, where the payload is PB-$i$ = TB-$i$ $\oplus$ PRNG-$i$. When sending a ``0'', the sender does nothing, but the receiver needs to call the synchronization function (time $t_b$). When sending a ``1'', the sender first performs the synchronous operation (time $t_b$), and then the receiver also calls the synchronization function (time $t_u$). When the number of ``0''s and ``1''s are equal, the average time for the sender to transmit a bit ($t_b$ / 2) is less than one for the receiver to receive a bit (($t_b$+$t_u$) / 2). Therefore, we ensure that the sender is ahead of the receiver. To control the lead of the sender, we add a waiting time of $t_s$ when sending ``0'', where $t_s$ is slightly smaller than $t_u$. It significantly reduces the lead by the sender. However, this lead accumulates with the transmission of bits, which may become relatively large after a long period of time. Therefore, we set a synchronization period, such as resynchronizing after transmitting 20000 bits. In fact, the period depends on the number of files or pages used, i.e., the time required to wrap around a circle. Through this method, the transmission of multiple bits can be achieved, and it can be continuously looped.

\subsection{Optimization 3: One-time Transmission}
Optimization 2 allows for the efficient use of shared resources, as multiple bits can be transmitted using the same set of files. However, this mode of communication requires precise measurement and control of the running time of both the sender and receiver to ensure that the sender is ahead of the receiver. This has certain requirements on the transmitted information, because even random encoding may occur with consecutive bits of ``1'' or ``0''. {\attack} is a persistent channel that allows the sender and receiver to execute non-concurrently. Thus, we propose a third scheme--one-time transmission, in which each bit transmitted is marked by one file or page. The sender chooses to perform or not perform synchronization based on the secret data, and the receiver keeps performing synchronization and timing to infer the secret data. If sufficient resources are shared, all bits can be sent at once.

The one-time transmission separates the execution of the sender and receiver, eliminating the need for any synchronization between two communication entities. Moreover, it only incurs tiny interference because the shared files or pages are separately utilized by the sender and receiver processes. This method can loosen \textbf{R1} and \textbf{R2}, thereby improving the performance of our channel. However, this method is limited by the number of files and the periodic write-back mechanism \cite{b42} on the target OS. For transmission with multiple files, a process can only use a limited number of file descriptors, which is no more than 1024 on Linux. For transmission with multiple pages, only one shared file is required. The pages that \texttt{msync} synchronizes are derived from \texttt{mmap} mapped files. Notably, although \texttt{msync} is not tailored for page size, it behaves as synchronization to page boundaries. Therefore, we use one page to represent one bit. In this way, a 1MB-sized file can transmit 256-bit secret data. The periodic write-back mechanism refreshes the dirty pages that have not been written back for more than 30 seconds by default, so the interval between the sender and the receiver cannot exceed 30 seconds.

In summary, Optimization 1 can exponentially enhance the TR, but it also leads to an exponential growth in the number of threads and files utilized, with a significant increase in the BER. Optimization 2 can improve the TR and reliability, but it requires precise control over the gaps between the sender and receiver, which may be influenced by the data transmitted (e.g., consecutive ``1''s or ``0''s). Optimization 3 separates the sender and receiver, increasing the TR while reducing interference, but it demands a sufficient amount of shared resources to tag each sent bit. In contrast, Optimization 1 and Optimization 2 can cycle multiple bits over a set of (small) files during transmission. Thus, these optimization strategies can be chosen based on the actual attack situations.

\section{Communication Synchronization}

In order to coordinate communication between both entities, it is necessary to synchronize the execution between the Trojan and the Spy at the beginning of the communication. The synchronization we discuss refers to communication synchronization rather than memory-disk synchronization. Specifically, before transmitting the $m$-bit secret data, the Trojan sends an $n$-bit pre-negotiated bit stream (e.g., ``10101010''), called the ``synchronization sequence''. The Spy continuously executes synchronization and timing until it receives a ``1''. Then the Spy verifies whether the initial $n$ bits of the received data match the ``synchronization sequence''. If they do, the Spy proceeds to receive the subsequent $m$-bit confidential data. Otherwise, the Spy discards the received data and prepares for the next round of reception. At the end of the communication, when the Spy receives a complete $n$ + $m$ bits, the reception is completed. Subsequently, the Spy can write to an unused file to signal the Trojan that this round is over and the next round of transmission is ready. This is because the Trojan has been measuring the synchronization time of the unused file after sending a round of data.

Furthermore, for communication protocols using a single file or page, we need fine-grained synchronization for the transmission of each bit, i.e., inter-symbol synchronization. Since our channel is based on the state changes of shared resources, the receiver must measure the state before transmitting the next bit. If any bit does not match, the correctness of all subsequent bits cannot be guaranteed. Thus, we need to reserve the execution time for the receiver after the sender's completion, i.e., the slot setting. For protocols using multiple files or pages, we do not need to consider the inter-symbol synchronization, since ensuring the sender executes before the receiver is enough to allow both processes to run in an expected order.

\section{Evaluation}
The capability of a covert channel can be measured by the following metrics \cite{b43}. 1) \textit{Bandwidth / TR.} The higher the rate, the greater amount of sensitive data an attacker can divulge. 2) \textit{BER}. The lower the error rate, the easier it is to recover valid data. 3) \textit{Time resolution.} This metric denotes the time interval between transmitting a ``0'' and a ``1''. When this interval is large, the receiver can dispense with a high-resolution clock. 4) \textit{Retention time.} The metric indicates a duration time during which the channel can retain the secret data. In volatile channels, this duration is 0 because concurrent execution is necessary for the sender and receiver. 5) \textit{Sharing level.} The level of sharing determines the applicability of covert channels, which is influenced by whether the sender and receiver are co-located on the same core, across cores, or across CPUs. 6) \textit{Stealthiness.} This metric indicates whether the channel is easy to be detected. Hardware covert channels can be detected by performance counter \cite{b6}, but there is no general detection method for software covert channels due to the diversity of software resources.

In this section, we rigorously evaluated {\attack} across diverse scenarios, examining time resolution, transmission rate (TR), bit error rate (BER), standard deviation (SD), and standard error (SE). Moreover, we compared it with published work. Experimentally, we employed three diverse machines: a laptop equipped with an AMD Ryzen 5 3500U processor ($2.1$ GHz), 8GB DDR4, and a 512GB Samsung SSD, running Ubuntu 20.04 and macOS 12; a MacBook featuring an M1 chip, operating on macOS 13; and a desktop with an Intel i5-8400 processor and Ubuntu 16.04, used for hardware-agnostic validation. In the experimental setup, the Trojan and Spy programs operated on the target OS, sharing the same files or pages during transmission. To enhance the accuracy of experimental results, we conducted each test $50$ times, transferring $1024$ bits in each iteration to obtain the average values of TR and BER.

\begin{figure}[t]
\centering
\includegraphics[width=\linewidth]{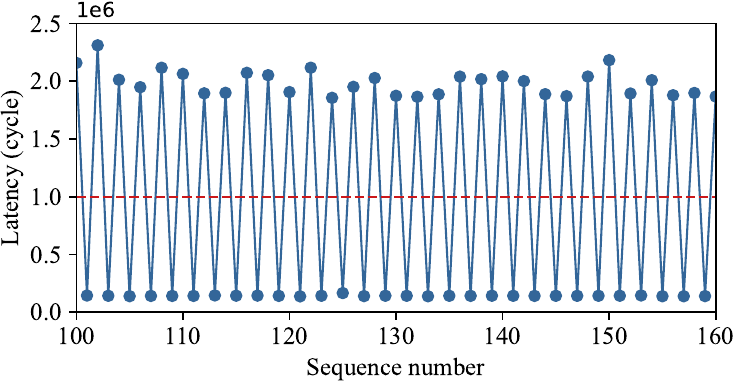}
\DeclareGraphicsExtensions.
\caption{The timing of sending ``0'' and sending ``1'' with \texttt{fdatasync} on Linux}
\label{fig4}
\end{figure}

\subsection{Experimental Results on Linux}
\textbf{Time resolution}. To demonstrate that the time difference between a buffer hit and miss is distinguishable, we first conducted an experimental verification by crafting a test process. We tested $1000$ sets of data, where the process sequentially chose to write or not write to the target file each time, and then measured the time it took to reclaim the write buffer via a synchronous system call. If the file is written, the reclaimed latency is higher. Otherwise, the latency will be lower. Fig. \ref{fig4} shows the difference in latency, which contains a portion of the tested data. We can see that the synchronization delay is about $0.918$ ms when the write buffer is cached and about 64 \textmu s when it is uncached. The time difference between the two cases is significant (over ten times), so we can set a threshold of $0.476$ ms ($1.0 \times 10^6$ cycles) to readily distinguish them.

\begin{figure}[t]
\centering
\includegraphics[width=\linewidth]{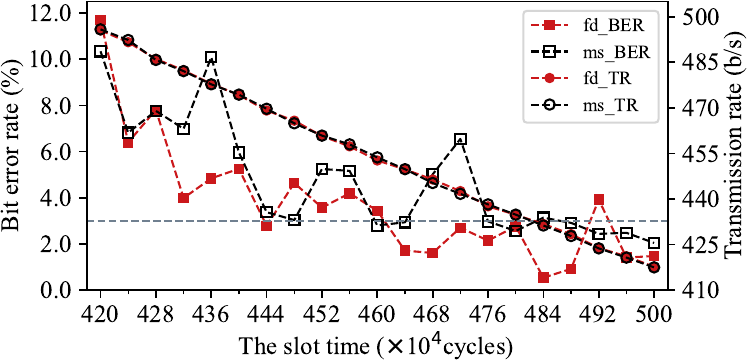}
\DeclareGraphicsExtensions.
\caption{The TR and BER of Protocol 1 and Protocol 2 (fd: fdatasync, ms: msync)}
\label{fig5}
\end{figure}

\begin{figure}[t]
\centering
\includegraphics[width=\linewidth]{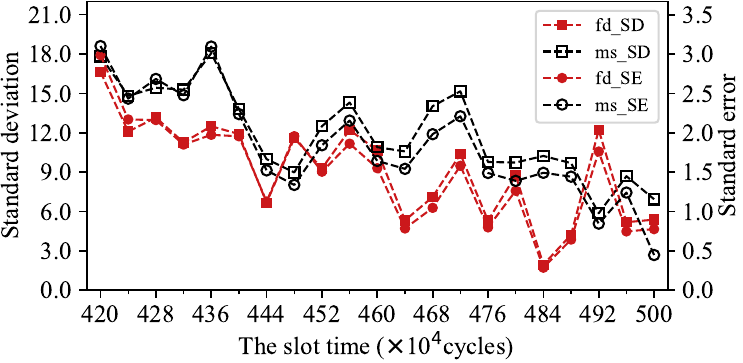}
\DeclareGraphicsExtensions.
\caption{The SD and SE of Protocol 1 and Protocol 2}
\label{fig6}
\end{figure}

\textbf{Single file or page.} We conducted experiments using \texttt{fdatasync} and \texttt{msync} for a single file and page, respectively. In this case, one slot for each transferred bit is required, the size of which can be adjusted depending on the trade-off between BER and TR. As shown in Fig. \ref{fig5}, we observed a significant decrease in the BER and a linear drop in the TR as the slot size increased. We can see that the BER drops below 3\% when the TR is about $430$ b/s. In addition, even though at a higher BER (e.g., slot $436$), we observed multiple sets of completely correct data. This indicates that we can run the tests multiple times to recover the secret data under such a circumstance. This observation is also consistent with Fig. \ref{fig6}, where we can see that the SD is large because there are multiple sets of perfectly correct and multi-bit wrong data. In addition, we find that the SE is small, indicating that our sample is relatively representative.

\textbf{Multiple files or pages.} In this case, we tested three optimization strategies: multi-bit encoding, asynchronous communication, and one-time transmission, respectively.

In multi-bit encoding, we used four files and four threads to transfer two bits simultaneously. Experimental results show that the BER increases significantly when the TR increases to $698.47$ b/s. In particular, when measuring the synchronization time in multiple threads, the execution time of the threads that are ahead of each other is shorter due to limited system resources. This behavior may affect the execution time of synchronous system calls, which is appreciable in our tests because our processor contains only 4 cores. However, we believe that the interference can be further reduced by improvements in hardware and code.

\begin{figure}[t]
\centering
\includegraphics[width=\linewidth]{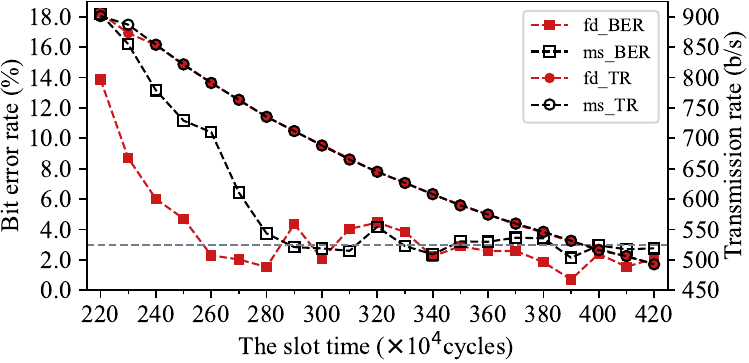}
\DeclareGraphicsExtensions.
\caption{The TR and BER of asynchronous communication with slot}
\label{fig7}
\end{figure}

\begin{figure}[t]
\centering
\includegraphics[width=\linewidth]{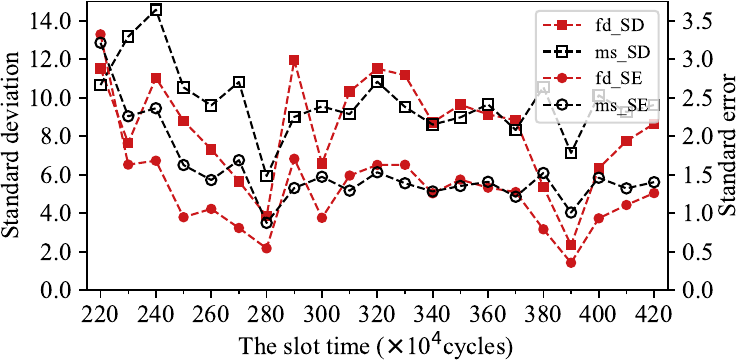}
\DeclareGraphicsExtensions.
\caption{The SD and SE of asynchronous communication}
\label{fig8}
\end{figure}

In asynchronous communication, we first retained the slot. In this case, the sender leads the receiver by one slot time, which would be maintained throughout the transmission. The receiver can continuously execute the synchronization function and measure its time without waiting for the execution of the sender. We tested this method using $8$ files and pages, and the corresponding results are shown in Fig. \ref{fig7}. We observed that the BER dropped to below $3$\% as the slot gradually increased, while the TR of our channel exceeded $700$ b/s. Fig. \ref{fig8} illustrates the SD and SE, which are similar to those of a single file or page. The large SD is attributed to the inclusion of multiple sets of entirely correct and multiple-bit erroneous data, influenced by the slot size and interference from synchronous system calls. We also evaluated Optimization 2 without a slot. In this case, we carefully designed the encoding method between the sender and receiver to ensure that the sender always led the receiver by a confined gap. To make the synchronous system call stable and the period of synchronization large, we used $500$ files for asynchronous communication without a slot. In our experiment, when $t_s$ is set to $0$, the communication needs to be resynchronized for every $1,600$ bits transmitted; when $t_s$ is set to $20$ \textmu s, it can be resynchronized after $20,000$ bits. Moreover, the initial sleep time of the receiver is set to $3000$ \textmu s. The asynchronous communication method without a slot further improves the TR but results in a higher BER. These errors occur especially in the early stages of the transmission process, in which the time of synchronous system call is not stable enough. In Table \ref{tab2}, we observed a TR of up to $1.625$ Kb/s and a BER of $6.5$\%.

In one-time transmission, we tested the TR and BER with multiple files and pages. When using multiple files, we can transmit $1022$ bits at a time due to the limitation of the number of files that a process can open. With multiple pages, we can easily transmit $6400$ bits using a $25$ MB file at once. Since this method is affected by the periodic write-back mechanism, which refreshes dirty pages not written back for over 30 seconds by default, it is advisable to ensure that the receiver lags behind the sender by no more than $20$ seconds. Moreover, in $50$ sets of data we tested ($1024$ bits for per set), we found no bit errors using this approach. Depending on the proportion of ``0''s and ``1''s in the sent data, we achieved an average TR of $2.036$ Kbps and a peak TR of $14.762$ Kbps. 

The performance of alternative synchronization functions is summarized in Table \ref{tab2}. Notably, \texttt{fsync} and \texttt{sync} exhibit slower TRs as they not only synchronize file data but also synchronize all file metadata to the disk. Due to the disparate storage of file data and metadata, these functions require additional time to complete the synchronization process.


\subsection{Experimental Results on Other OSes}
To illustrate the potential risks of {\attack}, we conducted experiments on other POSIX-compatible OSes. Firstly, we installed macOS Monterey 12 on an AMD processor and tested system calls such as \texttt{fsync}, \texttt{fdatasync}, and \texttt{msync}. The experimental results revealed that all these system calls successfully implemented a covert channel. Intriguingly, the call time for \texttt{fdatasync} and \texttt{msync} on this system is considerably shorter than on Ubuntu.

Taking \texttt{fdatasync} as an example, when the write buffer is cached, the synchronization time is approximately $0.176$ ms, whereas when it is not cached, the synchronization time is only about $3.6$ \textmu s. This enables {\attack} to achieve a TR far beyond that on Ubuntu. We demonstrated this extremely high transfer rate using Optimization 3, as shown in Table \ref{tab3}. The average rate signifies an equal distribution of ``0'' and ``1'', each occupying 50\% during data transmission. Meanwhile, the peak rate exemplifies a scenario where ``1'' exclusively occupies 100\%. Thus, contingent upon the transmitted data, our channel's speed can be notably enhanced. Next, we performed tests on a physical machine equipped with an M1 chip, running macOS Ventura 13. In this OS, we found that \texttt{fdatasync} and \texttt{msync} were not supported, although \texttt{fsync} remained functional. However, when testing with \texttt{fsync}, no observable time differences were noted. We speculate that macOS might ignore the disk-write requirements associated with \texttt{fsync}, failing to physically write the file data to the disk. This operation poses potential security risks, as data that should have been written to the disk may be lost in case of power failure or forced reboot. Additionally, even if the original POSIX system calls are not applicable on macOS, another function \texttt{fcntl} can still be used with \texttt{F\_FULLFSYNC} as a parameter to write data to permanent storage forcibly. Using this function, we achieved a TR of 117.25 b/s on macOS 13 based on a single shared file.

\begin{table}[tbp]
\centering
\small
\renewcommand{\arraystretch}{1.0}
\setlength{\abovecaptionskip}{0.2cm}
\caption{The TR and BER on Linux}
\label{tab2}
\begin{threeparttable}
\begin{tabular}{m{1.0cm}<{\centering} | m{3.5cm}<{\centering} | m{1.25cm}<{\centering} | m{1.3cm}<{\centering}}
\hline
\textbf{Type}  & \textbf{Strategy} & \textbf{TR (bps)}  & \textbf{BER (\%)}  \\
\hline
\multirow{4}{*}{\makecell[c]{Single}} & \texttt{sync} &  116.28    &  0.73   \\

 & \texttt{fsync}	& 139.66 	& 0.94 \\
& \texttt{fdatasync} &  469.12   &  2.79   \\
& \texttt{msync} &  453.72   &  2.81   \\
                             
\hline
\multirow{6}{*}{\makecell[c]{Multiple}} &  Optimization 1  & 698.47 & 5.56 \\
&  Optimization 2\_slot (file)  & 791.68 &  2.29 \\
&  Optimization 2\_slot (page)  & 711.86 &  2.84 \\
&  Optimization 2 (file)   & 1625.76 &  6.50 \\
&  Optimization 3 (file)   & 2013.54 &  0 \\
&  Optimization 3 (page)   & 2036.68 &  0 \\
\hline
\end{tabular}
\end{threeparttable}
\end{table}

\begin{table}[t]
\small
\renewcommand{\arraystretch}{1.0}
\setlength{\abovecaptionskip}{0.2cm}
\caption{The TR and BER on macOS}
\centering
  \begin{tabular}{cccc}
    \hline
    \textbf{Strategy} & \textbf{Ratio (\%)} &\textbf{TR (bps)} & \textbf{BER (\%)} \\
    \hline
    \multirow{3}{*}{\makecell[c]{Optimization 3 \\(file)}}
    & 50 & 9957.11 & 0 \\
   & 75 & 18569.22 & 0.0081 \\
   & 100 & 231392.37 (Peak) & 0 \\
   \hline
   \multirow{3}{*}{\makecell[c]{Optimization 3 \\(page)}} & 50 & 10211.41 & 0.004 \\
   & 75 & 19028.50 & 0.003 \\
   & 100 & 253021.92 (Peak) & 0 \\
    \hline
  \end{tabular}
\label{tab3}
\end{table}

\begin{table}[t]
\small
\renewcommand{\arraystretch}{1.0}
\setlength{\abovecaptionskip}{0.2cm}
\caption{The TR and BER in sandbox scenarios}
\centering
  \begin{tabular}{ccc}
    \hline
    \textbf{Strategy} & \textbf{TR (bps)} & \textbf{BER (\%)} \\
    \hline
   Optimization 3 (file) & 2003.45 (Peak: 13779) & 0.0058 \\
   Optimization 3 (page) & 2013.36 (Peak: 13830) & 0.0086 \\
    \hline
  \end{tabular}
\label{tab4}
\end{table}

The Windows system isn't POSIX-friendly by default, but we brought in \texttt{Cygwin} to bridge the gap. Surprisingly, there were no leaks in our experiments. We also tried Windows' own \texttt{FlushFileBuffers} and \texttt{FlushViewOfFile} functions for a similar transmission, but no luck there. This may be because: 1) these synchronization functions can only be used for files with write permissions; 2) synchronization functions between different processes may contain permission checks, or the write buffer of different processes may not maintain a consistent view. These functions can only refresh the file view of the current process, rather than affecting the view of other processes; 3) Each process has its own working set, and the write buffer may be in the working set \cite{b32}.

\subsection{Attacks on Cross-sandbox Scenarios}
We employed Firejail \cite{b37}, a commonly used open-source tool that offers a software sandbox environment for Linux, to evaluate the performance of {\attack}. In the Firejail sandbox, we blocked all network traffic and prohibited writing to memory, thereby preventing inter-process communication through shared memory. Firejail supports memory mapping using \texttt{mmap}, which allows us to call \texttt{msync} effortlessly. Therefore, we successfully implemented a write covert channel by synchronizing pages in the cross-sandbox scenario. Additionally, we enforced read-only access to all files in the sandbox using the command of ``read-only/HOME'', which restricted the Trojan from communicating directly with the Spy through shared files. Then, we redo the proposed covert channel using multiple files. Table \ref{tab4} shows that our covert channel is almost unaffected in the sandbox scenario, with a slightly reduced TR and an extremely low BER. 

\subsection{Comparison}


\subsubsection{{\attack} vs. Software Covert Channels}
Software covert channels do not rely on specific CPU components, so they can naturally communicate across CPUs, but only a few works have been proposed. These works can be classified into two categories \cite{b47}: volatile and persistent channels. In volatile channels, MES-Attacks \cite{b33} utilized mutual exclusion and synchronization mechanisms to implement a group of covert channels, achieving a TR of 7.182 Kb/s on Linux and a maximum TR of 13.105 Kb/s on Windows. Gao et al. \cite{b20} analyzed leakage channels in container cloud environments and proposed lock-based and memory-based channels, which can achieve 5150 b/s with 8 locks and 13.603 b/s with allocating 65\% of bit 1 and 35\% of bit 0 on Linux, respectively. The TR of our channel competes with that of these channels, but our channel does not depend on concurrent execution. A similar work synchronized with this paper is Sync+Sync \cite{b51} proposed by Jiang et al. However, the two focus on different vulnerability exploits. Sync+Sync discovered that the shared global log and serialization of commit transactions cause \texttt{fsync} to incur a longer response time during concurrency, and thus exploits this concurrent contention to enable covert channel and side-channel attacks. In contrast, the vulnerability we found is that synchronization system calls can be used in read-only mode. Based on this, we exploit the time difference generated by synchronizing cached and uncached data to disk to achieve persistent channels.

In persistent channels, Goethem et al. \cite{b12} proposed a covert channel based on browser cache. Although they do not provide a TR, the time resolution for accessing the cache ranges from a few milliseconds to tens of milliseconds. Therefore, our channel has a superior TR compared to theirs. Gruss et al. \cite{b9} introduced a page cache attack, which utilizes the \texttt{mincore} function to measure whether a memory page is present in the page cache, thereby leaking the accesses of sensitive pages. They also provided a covert channel with a TR of 56.32 Kb/s on Linux. However, the \texttt{mincore} function has been patched \cite{b36} and no longer provides accurate information about the page cache. We verified that this channel does not work on our systems. Overall, our channel is currently exploitable and has a high TR.

\subsubsection{{\attack} vs. Hardware Covert Channels.}
Currently, many hardware covert channels are proposed. Gruss et al. \cite{b3} demonstrated that Flush + Flush can achieve bit-rates of $496$ KB/s at less than 1\% BER. Using the last-level cache, Flush + Reload \cite{b2} achieved a capacity of $298$ KB/s, while Prime + Probe \cite{b4} implemented a TR of $75$ KB/s. Streamline \cite{b5} achieved an optimal bit rate of $1801$ KB/s at 0.37\% BER; Wu et al. \cite{b18} proposed a memory bus based channel with a capacity of $746$ b/s; DRAMA \cite{b14} achieved a capacity of $1200$ Kb/s on row buffers with a BER of 0.4\%. 

In general, it is unfair to compare the TR of software and hardware covert channels  due to the nature of the hardware itself. Despite the lower TR, {\attack} offers the following advantages: 1) it does not require specific hardware and complex reverse engineering; 2) it cannot be alleviated at present. Hardware cache channels can typically be mitigated by cache partitioning and randomization, while software cache channels can only be mitigated through targeted measures. Since software resources are numerous and diverse, complete isolation is impractical; 3) it is easier to implement. {\attack} does not rely on specific instructions, cache eviction sets, and cache replacement strategies. It can be implemented with simple system calls.

\section{Practical Attacks}
To show the security impact of {\attack}, we use it to demonstrate two practical attacks.

\textbf{Website Fingerprinting:} We first designed a website fingerprinting attack using {\attack}, in which the attacker utilizes the synchronization time of \texttt{fsync} to distinguish between different websites that the user is browsing. Initially, we randomly selected 20 websites from the same list as Rimmer et al. \cite{b50} and prepared an attacker process (Proc$_{A}$) and a victim process (Proc$_{V}$). 
The Proc$_{V}$ simulates the user's browsing behavior by periodically opening and closing websites (one tab at a time). The Proc$_{A}$ continuously uses \texttt{fsync} to synchronize files related to website I/O (such as \texttt{cookies.sqlite-wal} and \texttt{places.sqlite-wal}) and monitors the write operations of these files. The \texttt{WAL} file itself does not contain user privacy information. It is typically a logging mechanism used by database systems during transaction execution to ensure data integrity and consistency. Different websites may exhibit different I/O behaviors, leading to differences in the traces of synchronization time on these files. We conducted 100 tests for each website, collecting 100,000 traces each time, to generate our dataset. Among them, 70\% of the data is used as a training set and the remaining 30\% is used as a test set. Browser caches are not cleared before accessing each website. 

Subsequently, we employed the CNN algorithm \cite{b53} to classify these websites. Compared with Sync+Sync \cite{b51}, we focus on the number of \texttt{writes} rather than the number of \texttt{fsyncs} on each website. Additionally, the delay difference caused by \texttt{fsync} contention is smaller than that caused by synchronizing written data. Thus, we can obtain traces related to \texttt{writes} from the data set, thereby eliminating the effects of \texttt{fsync} contention. Table \ref{tab5} shows the classification accuracy, F1 score and the average number of \texttt{write} calls for each website. Although it may not represent the actual number of \texttt{writes}, it reflects the high latency count caused by \texttt{writes} during our testing period. Most websites exhibit similar numbers of write operations, but some also demonstrate distinct behaviors, enabling {\attack} to identify them with high accuracy. For instance, the accuracy rates for \texttt{bukalapak.com} and \texttt{archive.org} are 96.3\% and 86.4\%, respectively. Therefore, our attack can distinguish between websites with frequent or infrequent I/O operations.

\begin{table}[t]
\tiny
  \centering
  \caption{Website Classification of {\attack}}
  \setlength{\tabcolsep}{4.5pt} 
  \begin{subtable}[t]{0.46\linewidth}
    \centering
    \begin{tabular}{p{28pt}p{16pt}p{17pt}p{20pt}}
      \toprule
      Website & \texttt{\#Write} & Accuracy & F1-score \\
      \midrule
      imdb.com & 22.46 & 33.3\% & 0.50 \\
      360.cn  & 14.43  & 71.4\%  & 0.83 \\
      adobe.com & 22.95 & 69.7\% & 0.82 \\
      airbnb.com  & 25.83  & 36.1\%  & 0.53 \\
      abs-cbn.com  & 18.82  & 86.1\%  & 0.93 \\
      allegro.pl & 25.02 & 11.1\% & 0.2 \\
      amazon.com  & 19.48  & 37.5\%  & 0.55 \\
      apple.com & 20.55 & 38.7\% & 0.56 \\
      \textbf{archive.org}  & \textbf{10.36}  & \textbf{86.4\%}  & \textbf{0.93} \\
      baidu.com & 20.20 & 34.4\% & 0.51 \\
      \bottomrule
    \end{tabular}
  \end{subtable}
  \hfill
  \begin{subtable}[t]{0.48\linewidth}
    \centering
    \begin{tabular}{p{30pt}p{16pt}p{17pt}p{20pt}}
      \toprule
      Website & \texttt{\#Write} & Accuracy & F1-score \\
      \midrule
      bing.com  & 22.43  & 52.0\%  & 0.68 \\
      booking.com & 23.96  & 8.3\%  & 0.15 \\
      \textbf{bukalapak.com}  & \textbf{42.63}  & \textbf{96.3\%} & \textbf{0.98} \\
      canva.com & 31.70 & 43.3\% & 0.60 \\
      csdn.net  & 22.57  & 28.1\%  & 0.44 \\
      ebay.com & 19.22 & 66.7\% & 0.80 \\
      espn.com  & 26.61  & 41.7\%  & 0.59 \\
      espncricinfo.com & 23.49 & 14.3\% & 0.25 \\
      zoom.us  & 22.23  & 50.0\%  & 0.67 \\
      zhanqi.tv & 19.76 & 42.1\% & 0.59 \\
      \bottomrule
    \end{tabular}
  \end{subtable}
\label{tab5}
\end{table}

\textbf{Performance Degradation:} An attacker can repeatedly invoke the synchronization operation to flush the data written by the victim to disk, resulting in a significant increase in CPU wait time for the victim. Since the victim needs to continuously fetch data from disk instead of the page cache. This can degrade the performance of the victim's workload, thus launching some practical attacks such as a Dos attack \cite{b54} or amplification of side-channel leakage \cite{b55}. To test the effectiveness of this attack, we ran two processes pinned to different cores. Process 1 constantly invokes synchronous system calls, but does not generate any I/O operations itself. Process 2 runs \texttt{fio} benchmark \cite{b52} tests to measure I/O performance for five different workloads: sequential/random writes, sequential/random reads, and random reads and writes. The baseline is the bandwidth of process 2 when process 1 is not running. Moreover, the performance degradation is calculated by dividing the bandwidth of process 2 after enabling process 1 by the baseline. Experimental results show that with process 2 turned on, the benchmark performance is 61.07\%, 71.52\%, 44.73\%, 87.55\% and 43.86\%, respectively. Therefore, an attacker can implement a performance degradation attack by synchronizing functions.

\section{Discussion}
In this section, we will discuss ways to improve the stealth capabilities of the { \attack } and explore three potential mitigation strategies.

\subsection{Stealthiness}
In addition to TR and BER, stealthiness is a crucial factor to consider when designing a channel. Currently, there is no general detection method for software covert channels due to the diversity of software resources. However,  it is still possible to detect attacks by observing abnormal behavior in the system. For example, {\attack} might be detected by analyzing disk write rates or \texttt{fsync} call patterns. However, to reduce the observability of covert channel activity, an attacker can fool the detector by analyzing the function calls of a benign program and simulating its behavior with an adaptive version. For example, a fixed-length periodic transmission can be easily detected. To address this, the attacker can consider using variable-length transmission methods, where the sender and receiver negotiate in advance to use $n$ bits to represent the length of data to be transmitted each time. Since the number of bits transmitted is different each time, the disk write rate is not fixed. In addition, the intervals of transmission can also be adjusted to match the behavior of benign workloads to avoid detection. Although this approach reduces the rate of the channel, it brings greater flexibility and higher concealment. More schemes to improve stealthiness are also worth exploring in the future.

\subsection{Mitigation Strategies}
{\attack} exploits the delayed write mechanism and memory-disk synchronization primitives. Obviously, disabling the delayed write can block our channels completely, but it will result in significant performance degradation. There are three main mitigation strategies that might be used to restrict {\attack}:  detection, patching, or isolation, which are discussed in detail below.

\textbf{1) Detection.} 
Hardware cache covert channels can often be detected through performance counters in commercial processors because cache hit/miss rates or distributions differ from normal programs. There are currently no specialized detection tools for software cache covert channels, but abnormal patterns in the system may be analyzed to infer them. For example, monitoring the calls of system functions or abnormal behaviors. \texttt{iotop} is a potentially effective tool that can monitor disk I/O usage of processes in real-time, thus possibly detecting {\attack}. However, as mentioned above, attackers may deceive detectors by emulating benign application behavior using adaptive versions. Therefore, detection-based methods are not foolproof mitigation strategies.

\textbf{2) Patching.} 
The synchronization functions can be invoked among different processes, even by an process with the read-only permission. To mitigate this misuse, we recommend that the development community promptly patch these system calls. Potential solutions include adding permissions or noise. The former involves granting privileges or implementing write access checks for these functions. The usage of privileges may impact normal program execution since regular users also need these functions to ensure secure write-backs. However, we argue that write access restrictions are essential for these functions. This solution incurs relatively low overhead, but even with write access permissions in place, attackers may still exploit the contention behavior of these functions \cite{b51}. The latter involves introducing random or constant return times for the functions, making it difficult for attackers to discern cache status by timing these function calls, thus impeding accurate covert transmission. Additionally, Sync+Sync \cite{b51} is also sensitive to noise. However, changing delays to synchronization functions may incur some performance overhead.

\textbf{3) Isolation.} 
The sharing of kernel write buffers among different processes can lead to one process evicting dirty pages belonging to another. To mitigate this issue, an effective approach involves isolating the kernel write buffers of different processes, akin to cache partitioning \cite{b10}. Specifically, when a process accesses a file, the kernel creates a separate file view for that process and maps relevant data pages into the process's virtual address space. When using a synchronization function, only the write buffer of the calling process can be flushed, without affecting other processes. This isolation can provide adequate security but may result in significant system overhead, especially when dealing with numerous processes. Furthermore, maintaining the write buffer can be a laborious task, and its scalability may face challenges.

\section{Related Work}
Covert channels can be categorized into software-based and hardware-based channels. 

\textbf{Software-based channels.} Compared with hardware-based channels, software-based channels are simpler to implement as they do not require reverse engineering of component structures. Container \cite{b20} transfers information by reading dynamic identifiers and performance data from memory-based pseudo-files, such as \texttt{procfs} and \texttt{sysfs}. MES-Attacks \cite{b33} utilize inter-process mutual exclusion and synchronization mechanisms to transmit secrets by adjusting the time of locking and condition waiting. Page cache \cite{b9} measures whether a page is in the page cache or not by using a \texttt{mincore} system call. Other covert channels make use of different software resources or techniques, such as browser cache \cite{b12} and memory deduplication \cite{b19}. Due to the abundance of software resources, the resulting covert channels remain worthy of investigation. Moreover, how to detect such attacks also stands as a crucial task for future endeavors.

\textbf{Hardware-based channels.} Various microarchitectural components can be used to construct hardware-based covert channels, among which cache covert channels are the most widely used, such as Flush + Reload \cite{b2}, Flush + Flush \cite{b3}, and Prime + Probe \cite{b4}. Additionally, some covert channels that utilize hardware resources outside the CPU have emerged, such as DRAM Row Buffer \cite{b14}, Memory Bus \cite{b18}, CPU mesh \cite{b39}, and Network-on-Chip \cite{b34}. They enable cross-core or even cross-CPU communication. Furthermore, new variants of cache covert channels have attracted much more attention. In pioneering work, literature \cite{b5} proposed the first cache covert channel based on asynchronous communication, which transmits each bit by accessing different addresses of a large shared array. WB \cite{b7} implemented the first write-back cache covert channel, which utilizes the cache replacement latency between a dirty and a clean cache line. Reload + Refresh \cite{b6} cleverly tracks the victim's cache accesses by using cache replacement strategies, without evicting the victim's cache line by force. It can bypass side-channel detection using performance counters. Prime + Scope \cite{b8} uses a specific prime access pattern to implement the observation on a single cache line, thereby optimizing the resolution of attacks.

\section{Conclusion}
This paper presents {\attack}, a new covert communication method that exploits memory-disk synchronization primitives. The proposed covert channels are achieved in POSIX-compliant OSes, offering high flexibility and applicability without additional hardware support. We designed two covert channel protocols and three optimized methods to demonstrate the effectiveness of {\attack} attack, and evaluate its performance on different OSes and in both local and sandbox settings. The results show that our channels achieve an average rate of $2.036$ Kb/s with a measured error rate of $0$\% on Linux and a high average rate of $10.211$ Kb/s with an error rate of 0.004\% on macOS. In addition, our channel can operate within a sandbox environment with negligible impact on performance. Our method redefines the threat model of write covert channels, providing a new perspective and implementation method for such channels. Ultimately, we demonstrate how to extend our channel to two practical attacks and propose three potential countermeasures. Future work can explore further enhancements to the stealthiness, speed and reliability of the channel, as well as its application to more practical attacks. We anticipate that this work will stimulate more attention and investigation into covert communication in future security system designs.

\bibliographystyle{IEEEtran}
\bibliography{main}

\begin{IEEEbiography}
[{\includegraphics[width=1in,height=1.25in,clip,keepaspectratio]{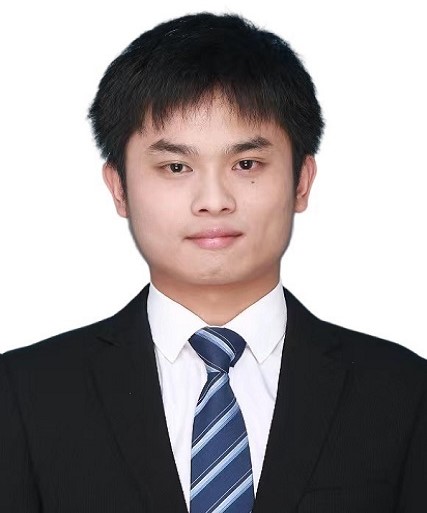}}]
{Congcong Chen} is currently pursuing the Ph.D. degree with Hunan University, China. His current research interests include architecture and microarchitecture security.
\end{IEEEbiography}

\begin{IEEEbiography}
[{\includegraphics[width=1in,height=1.25in,clip,keepaspectratio]{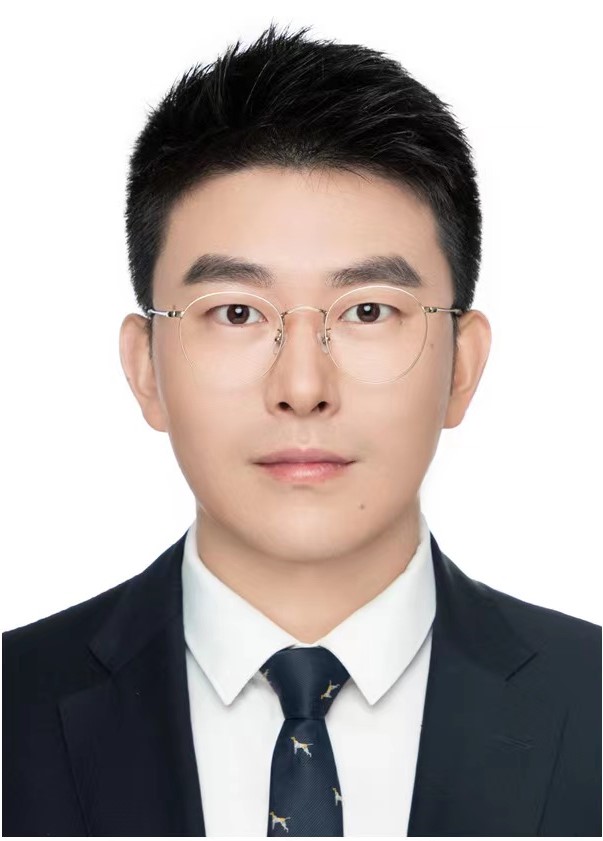}}]
{Jinhua Cui} is currently an Assistant Professor at Hunan University. He received the Ph.D. degree from National University of Defense Technology, Changsha, China, in June 2022. From July 2019 to September 2021, he was a Research Assistant at the School of Computing in National University of Singapore. Before that, he was also a Senior Research Engineer at the Secure Mobile Centre in Singapore Management University. His current research interests include trusted computing, microarchitectural security, and hardware-assisted system security.

\end{IEEEbiography}

\begin{IEEEbiography}
[{\includegraphics[width=1in,height=1.25in,clip,keepaspectratio]{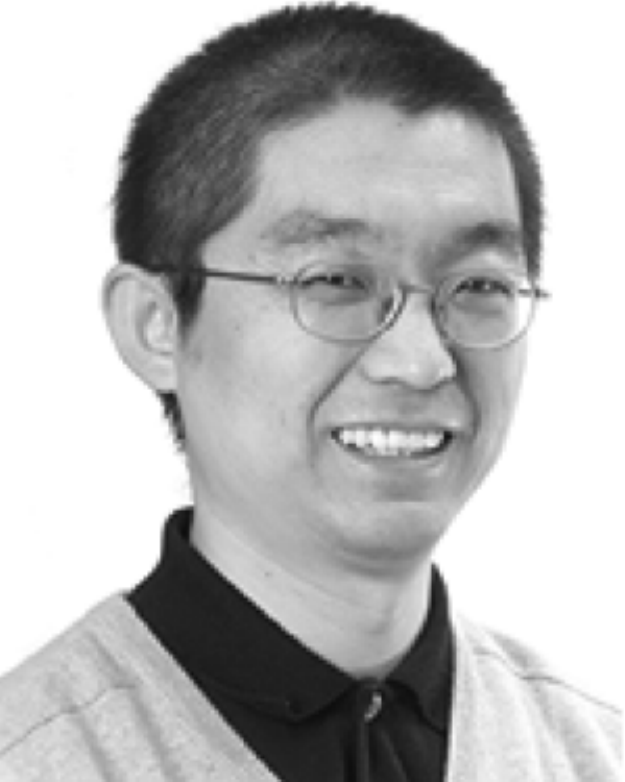}}]
{Gang Qu}
(Fellow, IEEE) received the B.S. and M.S. degrees in mathematics from the University of Science and Technology of China, in 1992 and 1994, respectively, and the Ph.D. degree in computer science from the University of California, Los Angeles, in 2000. Upon graduation, he joined the University of Maryland at College Park, where he is currently a professor in the Department of Electrical and Computer Engineering and Institute for Systems Research. He is also the director of Maryland Embedded Systems and Hardware Security Lab and the Wireless Sensors Laboratory.
His primary research interests are in the area of embedded systems and VLSI CAD with focus on low power system design and hardware related security and trust.  
\end{IEEEbiography}

\begin{IEEEbiography}
[{\includegraphics[width=1in,height=1.25in,clip,keepaspectratio]{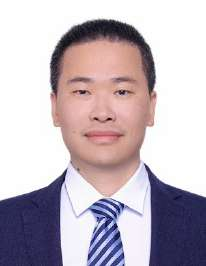}}]
{Jiliang Zhang}
(Senior member, IEEE) received the Ph.D. degree in Computer Science and Technology from Hunan University, Changsha, China in 2015. From 2013 to 2014, he worked as a Research Scholar at the Maryland Embedded Systems and Hardware Security Lab, University of Maryland, College Park. From 2015 to 2017, he was an Associate Professor with Northeastern University, China. In April 2017, he joined the Hunan University. He is currently a Full Professor at the College of Integrated Circuits, Hunan University. He is Vice Dean of the College of Integrated Circuits at Hunan University, the Director of Chip Security Institute of Hunan University, and the Director of CCF Fault-Tolerant Computing Professional Committee. His current research interests include Integrated Circuit Hardware Security, Low-power Integrated Circuit Design, Confidential Computing and so on. He has authored more than 80 technical papers in leading journals and conferences. He was the recipient of CCF Integrated Circuit Early Career Award, and CCF Distinguished Speaker, and the winner of Excellent Youth Fund of the National Natural Science Foundation of China. He has been the Program Committee Member for a number of well-known conferences such as DAC, ASP-DAC, GLSVLSI and FPT.
\end{IEEEbiography}

\vfill

\end{document}